\documentclass[11pt,pra,showkeys,showpacs]{revtex4}
\usepackage{amssymb,amsfonts,amsmath}
\usepackage{bbm,bm}
\newcommand{\p}{{\prime}}
\newcommand{\lz}{\lambda_0}
\newcommand{\lf}{\lambda_1}
\newcommand{\ls}{\lambda_2}
\newcommand{\lt}{\lambda_3}
\newcommand{\lh}{\lambda_4}
\newcommand{\ra}{\rangle}
\newcommand{\la}{\langle}

\newcommand{\ket}[1]{\vert #1 \ra}

\newcommand{\ov}[2]{\left\la #1 | #2 \right\ra}
\begin{document}

 \title{Nonstrict inequality for Schmidt coefficients of three-qubit states}

 \author{Levon Tamaryan}

 \affiliation{Theory Department, Alikhanyan National Laboratory(YerPhI), Yerevan, Armenia}

\begin{abstract}
Generalized Schmidt decomposition of pure three-qubit states has
four positive and one complex coefficients. In contrast to the
bipartite case, they are not arbitrary and the largest Schmidt
coefficient restricts severely other coefficients. We derive a
nonstrict inequality between three-qubit Schmidt coefficients,
where the largest coefficient defines the least upper bound for
the three nondiagonal coefficients or, equivalently, the three
nondiagonal coefficients together define the greatest lower bound
for the largest coefficient. In addition, we show the existence of
another inequality which should establish an upper bound for the
remaining Schmidt coefficient.
\end{abstract}



 \maketitle

\section{Introduction}

Tripartite entanglement is a difficult subject for physicists. Essential results were obtained in this field~\cite{hig,acin,coff,w}, but fundamental problems remain unsolved. Two of them are the main obstacles to understand tripartite entanglement so well as bipartite entanglement.

The first problem is the entanglement transformation problem. Its essence is the set of necessary and sufficient conditions for transforming a given pure tripartite state to another pure tripartite state by local operations and classical communication. This problem is solved for bipartite systems~\cite{niels} and therefore the conditions for bipartite entanglement transformation based on majorization give a concise answer to the questions: among given states which ones are more/less entangled and which ones are incomparable? Unfortunately these problem is a puzzle in the case of tripartite systems.

The second problem, closely related to the first one, is the
notion of maximally entangled states. This problem also is solved
for bipartite systems, and maximally entangled two-qubit states
are the  Einstein-Podolsky-Rosen state~\cite{epr} and its local
unitary (LU) equivalents known as  Bell states~\cite{bel}.
However, there is no clear and unique definition of a maximally
entangled state in multipartite settings. Consequently it is
impossible to introduce operational entanglement measures based on
optimal rates of conversion between arbitrary states and maximally
entangled states~\cite{ben,woot,dist-2}.

For bipartite systems these problems have been solved with the
help of the Schmidt decomposition~\cite{smid,ekert}. Therefore its
generalization to multipartite states can solve difficult problems
related to multipartite entanglement. This generalization for
three qubits is done by Ac\'in {\it et al.}~\cite{acin}, where it
is shown that an arbitrary pure state can be written as a linear
combination of five product states. Independently, Carteret {\it
et al.} developed a method for such a generalization for pure
states of an arbitrary multipartite system, where the dimensions
of the individual state spaces are finite but otherwise
arbitrary~\cite{hig}. The main idea of this method is the
following. First one finds the product vector which gives maximal
overlap with a given quantum state vector. Then one considers
product vectors orthogonal to the first product vector and finds
among them the product vector that gives maximal overlap with the
state vector. Continuing in this way, one finds a set of
orthogonal product states and presents the state function as a
linear combination of these product vectors. Since the first
product vector is a stationarity point, the resulting canonical
form contains a minimal set of state parameters.

Just as in the bipartite case, the largest coefficient of this
canonical form is the maximal product overlap which is an
increasing entanglement monotone~\cite{barn-01}. Just as in
bipartite case, the second largest coefficient is the maximal
overlap over product states orthogonal to the nearest product
state and so on. Additionally, this generalization of the Schmidt
decomposition(GSD) gives insight into the nature of the maximally
entangled three-qubit states~\cite{maxim} and is a good tool to
extend Nielsen's theorem and operational entanglement measures to
multipartite cases. Hence we accept that the amplitudes of the GSD
proposed in~\cite{hig} are multipartite Schmidt coefficients.

However, for a given quantum state the canonical form is not
unique and the same state can have different canonical forms and
therefore different sets of such amplitudes. The reason is that
the stationarity equations defining stationarity points are
nonlinear equations and in general have several solutions of
different types. For instance, three-parameter {\it W} type states
have four stationary points that create four equivalent canonical
forms for the same {\it W} type state. This point is explained in
detail in Sec. III, and now we focus on the question of which
amplitude sets should be treated as Schmidt coefficients and which
ones should be treated as insignificant mathematical solutions. A
criterion should exist that can distinguish right Schmidt
coefficients from false ones, and we need such a criterion. It is
unlikely that we can solve problems of three-qubit entanglement
without knowledge of what quantities are the relevant entanglement
parameters.

The canonical form whose largest coefficient is the maximal
product overlap, as in the bipartite case, presents the GSD, and
others are irrelevant solutions of stationarity equations. Then
our task is to single out the canonical form whose largest
coefficient is the maximal product overlap, and this requirement
gives rise to a nontrivial relation between Schmidt coefficients
of three-qubits. This situation differs from the bipartite case,
where each set of positive numbers satisfying the normalization
condition presents Schmidt coefficients of some quantum state and
its LU-equivalents. In contrast, in the three-qubit case four
positive and one complex coefficients satisfying the normalization
condition are Schmidt coefficients if they satisfy an equality
(derived in Sec. V), otherwise they do not present relevant
entanglement parameters at all. This is the main result of this
work.

It is clear how we single out the canonical form whose largest
coefficient is the maximal product overlap. We should single out
the closest product state of a given quantum state that gives a
true maximum for overlap.  Of course, we cannot find closest
product states of generic three-qubit states because there is no
method to solve a generic stationarity equation so far. Hence to
distinguish the true maximum from other stationary points we
require that the second variation of the maximal product overlap
be negative everywhere, and this condition yields the desired
inequality.

However, the derived nonstrict inequality is a necessary but not a
sufficient condition for specifying uniquely the Schmidt
coefficients. It establishes an upper bound for the three middle
coefficients, and this upper bound is defined by the largest
coefficient. But it does not give an upper bound for the last
coefficient, which also should have an upper bound conditioned by
four previous coefficients. The existence of an upper bound for
the last coefficient is clarified in Sec. IV which means that an
additional inequality is needed to distinguish clearly the right
Schmidt coefficients from the false ones.

This paper is organized as follows. In Sec. II we repeat the
derivation of the GSD for three-qubit systems. In Sec. III we
present an illustrative example showing that the canonical form is
not unique. In Sec. IV we compute the second variation of the
maximal product overlap. In Sec.V we derive the nonstrict
inequality for three-qubit Schmidt coefficients and analyze
particular cases. In Sec. VI we show that another inequality is
needed to specify uniquely Schmidt coefficients. In Sec. VII we
discuss our results.

\section{Generalized Schmidt decomposition for three-qubits}

In this section we derive GSD for three-qubit pure states in
detail since the derivation method is used in Sec. IV to compute
the second variation of the maximal product overlap.

For a three-qubit pure state $\ket{\psi}$ the maximal product overlap $\lz(\psi)$ is defined as
\begin{equation}\label{2.lzdef}
\lz(\psi)=\max|\la u_1u_2u_3|\psi\ra|,
\end{equation}
\noindent where the maximum is over all tuples of vectors
$|u_k\ra$ with $\|u_k\|=1, (k=1,2,3)$. Note, hereafter the labels
within each ket refer to qubits 1, 2, and 3 in that order.

To find the maximum of $\lz(\psi)$ with constraints $\|u_k\|=1$ we form the auxiliary function $\Lambda$ given by
\begin{equation}\label{2.Lzdef}
\Lambda=|\la u_1u_2u_3|\psi\ra|^2 + \alpha_1(\la u_1|u_1\ra-1 )+ \alpha_2(\la u_2|u_2\ra-1) +\alpha_3(\la u_3|u_3\ra-1),
\end{equation}
where the Lagrange multipliers $\alpha_k$ enforce the unit nature
of the local vectors $|u_k\ra$.

Now we consider small variations of $|u_k\ra$ and $\alpha_k$, that
is $|u_k\ra\to|u_k\ra+|\delta
u_k\ra;\,\alpha_k\to\alpha_k+\delta\alpha_k$, and compute the
resulting variation of $\Lambda$. Hereafter $\delta\Lambda$  and
$\delta^n\Lambda$ mean the full and the $n$th variation of
$\Lambda$, respectively.

First we consider the first variation and require that $\delta^1\Lambda=0$. Then the vanishing of the partial derivatives of $\Lambda$ with respect to the Lagrange multipliers $\alpha_k$ gives
\begin{equation}\label{2.real}
\la u_1|u_1\ra-1=\la u_2|u_2\ra-1=\la u_3|u_3\ra-1=0,
\end{equation}
which are constraints on the local states $\ket{u_i}$.

The vanishing of the partial derivatives of $\Lambda$ with respect to these local states gives
\begin{eqnarray}\label{2.stp}
  \nonumber
  \ov{\psi}{u_1u_2u_3}\ov{u_2u_3}{\psi}+\alpha_1\ket{u_1} &=& 0 \\
  \ov{\psi}{u_1u_2u_3}\ov{u_1u_3}{\psi}+\alpha_2\ket{u_2} &=& 0 \\
  \nonumber
  \ov{\psi}{u_1u_2u_3}\ov{u_1u_2}{\psi}+\alpha_3\ket{u_3} &=& 0
\end{eqnarray}
and their Hermitian conjugates. From \eqref{2.stp} it follows that $\alpha_1=\alpha_2=\alpha_3=-\lz^2$ and therefore we can adjust phases of $\ket{u_k}$ so that stationarity equations \eqref{2.stp} become
\begin{equation}\label{2.stf}
\ov{u_2u_3}{\psi}=\lz\ket{u_1},\quad\ov{u_1u_3}{\psi}=\lz\ket{u_2},\quad \ov{u_1u_2}{\psi}=\lz\ket{u_3}.
\end{equation}

In the case of three-qubit states these equations are sufficient
to construct the GSD as follows. For each single-qubit state
$|u_k\ra$ there is, up to an arbitrary phase, a unique
single-qubit state $|v_k\ra$ orthogonal to it. Then from
\eqref{2.stf} it follows that the product states
$$\ket{u_1u_2v_3},\quad\ket{u_1v_2u_3},\quad\ket{v_1u_2u_3}$$ are
orthogonal to $\ket{\psi}$ and \eqref{2.stf} can be written as
\begin{eqnarray}\label{2.stlup}
  \nonumber
  \ov{u_1}{\psi} &=& \lz\ket{u_2u_3}+\lf\ket{v_2v_3}, \\
  \ov{u_2}{\psi} &=& \lz\ket{u_1u_3}+\ls\ket{v_1v_3}, \\
  \nonumber
  \ov{u_3}{\psi} &=& \lz\ket{u_1u_2}+\lt\ket{v_1v_2}.
\end{eqnarray}
 We choose the phases of $\ket{v_k}$ such that $\lf,\ls,\lt\ge0$. Note that after this choice the collective sign-flip of the $\ket{v_k}$'s does not change anything, and we will use this freedom in a little while.

 The state $\psi$ can be written as a linear combination of five product states as follows
\begin{equation}\label{2.gsd}
\ket{\psi}=\lz\ket{u_1u_2u_3} + \lf\ket{u_1v_2v_3} + \ls\ket{v_1u_2v_3} + \lt\ket{v_1v_2u_3} + \lh\ket{v_1v_2v_3},
\end{equation}
where $\lh$ is a complex number. It has two constraints: first
$\lz\ge|\lh|$ and second $-\pi/2\leq {\rm Arg}(\lh) \leq\pi/2$,
which can be achieved by the simultaneous change of the signs of
the local states $\ket{v_k}$.

Sometimes one relabels $\ket{u}\to\ket{0},\;\ket{v}\to\ket{1}$ for
simplicity. We leave \eqref{2.gsd} as is and refer to it as the
GSD for three-qubits.

\section{Illustrative example}

In this section we show that for a given state $\ket{\psi}$ the
canonical form \eqref{2.gsd} is not unique except in rare cases,
and additional relations are needed to single out the Schmidt
decomposition from the useless canonical forms.

Consider a three-parameter family of {\it W} type states~\cite{w}
given by
\begin{equation}\label{3.w}
|w(a,b,c)\ra=a|100\ra+b|010\ra+c|001\ra,
\end{equation}
where parameters $a,b,c$ are all positive since their phases can be eliminated by appropriate LU transformations. Stationarity equations \eqref{2.stf} of this state have three simple solutions and one special solution which exists if and only if parameters $a,b,c$ can form a triangle~\cite{anal}.

The three simple solutions are
\begin{eqnarray}\label{3.soltr}
  \ket{u_1(1)}=\ket{1},\quad \ket{u_2(1)}=\ket{0},\quad\ket{u_3(1)}=\ket{0},\quad\lz(1)&=& a;\label{3.soltr1} \\
  \ket{u_1(2)}=\ket{0},\quad \ket{u_2(2)}=\ket{1},\quad\ket{u_3(2)}=\ket{0},\quad\lz(2)&=& b;\label{3.soltr2} \\
  \ket{u_1(3)}=\ket{0},\quad \ket{u_2(3)}=\ket{0},\quad\ket{u_3(3)}=\ket{1},\quad\lz(3)&=& c;\label{3.soltr3}
\end{eqnarray}
where numbers within parentheses mark solutions.

The fourth nontrivial solution is
\begin{eqnarray}\label{3.solnon}
  \nonumber
  \ket{u_1(4)} &=& \frac{a\sqrt{2r_a}\,\ket{0}+\sqrt{r_br_c}\,\ket{1}}{4S},\quad
  \ket{u_2(4)} \frac{b\sqrt{2r_b}\,\ket{0}+\sqrt{r_ar_c}\,\ket{1}}{4S} \\
  \ket{u_3(4)} &=& \frac{c\sqrt{2r_c}\,\ket{0}+\sqrt{r_ar_b}\,\ket{1}}{4S},\quad \lz(4)=\frac{abc}{2S}
\end{eqnarray}
where
\begin{equation}\label{3.bloch}
   r_a=b^2+c^2-a^2,\; r_b=a^2+c^2-b^2,\; r_c=a^2+b^2-c^2
\end{equation}
and $S$ is the area of the triangle $(a,b,c)$.

At $r_ar_br_c=0$ the special solution reduces to a trivial
solution. Note that absolute values of these quantities
$|r_a|,|r_b|,|r_c|$ are magnitudes of Bloch vectors of the first,
second and third qubits, respectively and $r_ar_br_c=0$ means that
it exists an one-particle reduced density which is a multiple of
the unit matrix. In other words, the states with a completely
mixed subsystems appear at the edge of the special solution and
viceversa.

These four solutions of \eqref{2.stf} give the following four canonical forms for the state \eqref{3.w}
\begin{eqnarray}\label{3.cf}
  \ket{w(a,b,c)} &=& \lz(1)\ket{u_1(1)u_2(1)u_3(1)} + b\ket{v_1(1)u_2(1)v_3(1)} + c\ket{v_1(1)v_2(1)u_3(1)},\label{3.cf1}\\
  \ket{w(a,b,c)} &=& \lz(2)\ket{u_1(2)u_2(2)u_3(2)} + c\ket{u_1(2)u_2(2)u_3(2)} +  a\ket{v_1(2)v_2(2)u_3(2)},\label{3.cf2}\\
  \ket{w(a,b,c)} &=& \lz(3)\ket{u_1(3)u_2(3)u_3(3)} + b\ket{u_1(2)v_2(2)v_3(2)} + a\ket{v_1(3)u_2(3)v_3(3)},\label{3.cf3}
\end{eqnarray}
\begin{eqnarray}\label{3.cf4}
 \nonumber
 \ket{w(a,b,c)} &=& \lz(4)\ket{u_1(4)u_2(4)u_3(4)} + \frac{ar_a}{4S}\ket{u_1(4)v_2(4)v_3(4)} + \frac{br_b}{4S}\ket{v_1(4)u_2(4)v_3(4)}\\
 &+& \frac{cr_c}{4S}\ket{u_1(4)u_2(4)v_3(4)} + i\frac{\sqrt{2r_ar_br_c}}{4S}\ket{v_1(4)v_2(4)v_3(4)}.
\end{eqnarray}

Now which of these canonical forms is a right decomposition?

It is easy to clarify this question in this particular case since we have all solutions of the stationarity equations \eqref{2.stf} and can single out the one whose largest coefficient is the dominant eigenvalue of \eqref{2.stf}.

The answer is~\cite{anal}:
\begin{enumerate}
  \item If $r_a<0$ then only $\lz(1)$ is the maximal eigenvalue of \eqref{2.stf}, but $\lz(2),\lz(3),\lz(4)$ are not.
  \item If $r_b<0$ then only $\lz(2)$ is the maximal eigenvalue of \eqref{2.stf}, but $\lz(1),\lz(3),\lz(4)$ are not.
  \item If $r_c<0$ then only $\lz(3)$ is the maximal eigenvalue of \eqref{2.stf}, but $\lz(1),\lz(2),\lz(4)$ are not.
  \item Otherwise only $\lz(4)$ is the maximal eigenvalue of \eqref{2.stf}, but $\lz(1),\lz(2),\lz(3)$ are not.
\end{enumerate}
However, we are unable to solve \eqref{2.stf} for generic states
and single out the maximal eigenvalue in this way. Also we are not
forced to compare all eigenvalues of \eqref{2.stf} to see whether
the largest coefficient of a given decomposition is the maximal
product overlap. We can just require that it is truly a maximum of
the product overlap instead and obtain a criterion which shows
whether the largest coefficient of a given canonical form is the
maximal product overlap of the state. This will be done in the
next sections.

\section{The second variation of the maximal product overlap}

In this section we compute the second variation of the maximal product overlap.

We compute it at stationary points to single out true maximuma,
and therefore we use the results coming from the vanishing of the
first variation. Straightforward calculation gives
\begin{eqnarray}\label{4.var1}
  \nonumber
  \delta^2\Lambda &=& \lz^2\left|\ov{\delta u_1}{u_1} + \ov{\delta u_2}{u_2} +\ov{\delta u_3}{u_3} \right|^2 - \lz^2\left(\|\delta u_1\|^2 +\|\delta u_2\|^2 +\|\delta u_3\|^2 \right)\\
  &+& \lz^2\left( \ov{\delta u_1}{u_1}\ov{\delta u_2}{u_2} + \ov{\delta u_1}{u_1}\ov{\delta u_3}{u_3} + \ov{\delta u_2}{u_2}\ov{\delta u_3}{u_3} + {\rm cc}\right)\\
  \nonumber
  &+&\lz\left(\lt\ov{\delta u_1}{v_1}\ov{\delta u_2}{v_2} + \ls\ov{\delta u_1}{v_1}\ov{\delta u_3}{v_3} + \lf\ov{\delta u_2}{v_2}\ov{\delta u_3}{v_3} + {\rm cc}\right)\\
  \nonumber
  &+&\delta\alpha_1\delta||u_1||^2 + \delta\alpha_2\delta||u_2||^2 + \delta\alpha_3\delta||u_3||^2,
\end{eqnarray}
where cc means complex conjugate.

Using the identity $\|\delta u_k\|^2\equiv\left|\ov{\delta u_k}{u_k}\right|^2 + \left|\ov{\delta u_k}{v_k}\right|^2 $ it can be rewritten as
\begin{eqnarray}\label{4.var2}
  \delta^2\Lambda &=& -\lz^2\left(\left|\ov{\delta u_1}{v_1}\right|^2 +  \left|\ov{\delta u_2}{v_2}\right|^2 +\left|\ov{\delta u_3}{v_3}\right|^2 \right) \\ \nonumber
   &+& \lz\left(\lt\ov{\delta u_1}{v_1}\ov{\delta u_2}{v_2} + \ls\ov{\delta u_1}{v_1}\ov{\delta u_3}{v_3} + \lf\ov{\delta u_2}{v_2}\ov{\delta u_3}{v_3} + {\rm c.c.}\right) \\ \nonumber
   &+& \lz^2\left(\delta||u_1||^2\delta||u_2||^2 + \delta||u_1||^2\delta||u_3||^2) + \delta||u_2||^2\delta||u_3||^2\right)\\ \nonumber
   &+&\delta\alpha_1\delta||u_1||^2 + \delta\alpha_2\delta||u_2||^2 + \delta\alpha_3\delta||u_3||^2.
\end{eqnarray}
From \eqref{2.real} it follows that terms containing $\delta||u_k||^2$ vanish and the second variation takes the form
\begin{eqnarray}\label{4.var3}
  \delta^2\Lambda &=& -\lz^2\left(\left|\ov{\delta u_1}{v_1}\right|^2 +  \left|\ov{\delta u_2}{v_2}\right|^2 +\left|\ov{\delta u_3}{v_3}\right|^2 \right) \\\nonumber
   &+& \lz\left(\lt\ov{\delta u_1}{v_1}\ov{\delta u_2}{v_2} + \ls\ov{\delta u_1}{v_1}\ov{\delta u_3}{v_3} + \lf\ov{\delta u_2}{v_2}\ov{\delta u_3}{v_3} + {\rm c.c.}\right).
\end{eqnarray}
From $\ov{\delta u_i}{v_i}\ov{\delta u_j}{v_j}\leq|\ov{\delta u_i}{v_i}\ov{\delta u_j}{v_j}|$ it follows that
\begin{equation}\label{4.est}
   \delta^2\Lambda\leq-\lz\sum_{i,j=1}^3|\ov{\delta u_i}{v_i}||\ov{\delta u_j}{v_j}|A_{ij},
\end{equation}
where the real and symmetric matrix $A$ is given by

\begin{equation}\label{4.mat}
A=
\begin{pmatrix}
\lz & -\lt & -\ls\\
-\lt & \lz & -\lf\\
-\ls & -\lf & \lz
\end{pmatrix}
.
\end{equation}
Note that the inequality \eqref{4.est} can be saturated when
vectors $\ket{\delta u_k}$ are all multiples of vectors $v_k$, and
therefore \eqref{4.est} gives the least upper bound of
$\delta^2\Lambda$.

\section{A non-strict inequality for the Schmidt coefficients}

In this section we derive a nonstrict inequality for the Schmidt
coefficients.

The condition $\delta^2\Lambda\leq0$ holds everywhere if and only
if the matrix $A$ is positive, which means that
\begin{equation}\label{5.mpos}
   {\rm tr}(A)\geq0,\quad [{\rm tr}(A)]^2- {\rm tr}(A^2)\geq0,\quad \det(A)\geq0,
\end{equation}
where tr and det mean the trace and the determinant of a matrix, respectively.

The first condition ${\rm tr}(A)=3\lz>0$  is satisfied and does
not give anything. Similarly, the second condition $[{\rm
tr}(A)]^2- {\rm tr}(A^2)=6\lz^2-2(\lf^2+\ls^2+\lt^2)>0$ is a
triviality since $\lz$ is the largest coefficient. But the third
condition $\det(A)\geq0$ gives
\begin{equation}\label{5.ineq}
   \lz^2\;\geq\;\lf^2+\ls^2+\lt^2+2\;\frac{\lf\ls\lt}{\lz}.
\end{equation}
This is a new and unexpected relation which says that nondiagonal
coefficients all together are bounded above by the quantity
depending only on the largest coefficient, and therefore they
should be small.

Let us consider some particular cases. First consider the case
when some nondiagonal coefficient, namely $\lf$, vanishes. Then
\eqref{5.ineq} reduces to
\begin{equation}\label{5.inl10}
   \lz^2\;\geq\;\ls^2+\lt^2,\quad\lf=0.
\end{equation}
 The solution \eqref{3.soltr1} and the canonical form \eqref {3.cf1} present this case. This happens when a quantum state is a linear combination of three product states and its amplitudes in a computational basis satisfy \eqref{5.inl10}. Then the largest amplitude is the largest Schmidt coefficient and the GSD is achieved by a simple flipping of local states. Similarly, the solution \eqref{3.soltr2} with the form \eqref {3.cf2} and solution \eqref{3.soltr3} with the form \eqref {3.cf3} are the cases $\ls=0$ and $\lt=0$, respectively.

 Conversely, when amplitudes of a three-term state in a computational basis do not satisfy \eqref{5.inl10}, there appears a special solution \eqref{3.solnon} which creates a new factorizable basis. In this basis new amplitudes of the state given by \eqref {3.cf4} satisfy \eqref{5.ineq}.  Indeed,
 \begin{equation}\label{5.intri}
 4(abc)^2\geq (ar_a)^2 + (br_b)^2 + (cr_c)^2 +r_ar_br_c,
 \end{equation}
 which can be checked using triangle inequalities. This means that if amplitudes of the state were not satisfying \eqref{5.ineq} in the initial basis from product states, then there appears a special solution giving rise to a new basis from product states, and in this final basis amplitudes do satisfy \eqref{5.ineq}.

 In conclusion, \eqref{5.ineq} clearly  indicates whether a given canonical form is a GSD or not and this is its main advantage.

 Another particular case which we would like to elucidate is the following. We want to find a quantum state for which \eqref{5.ineq} is saturated and nondiagonal coefficients have maximal values. We equate all nondiagonal coefficients for simplicity and \eqref{5.ineq} reduces to
\begin{equation}\label{5.insym}
   \lz\;\geq\;2\lambda,\quad\lf=\ls=\lt\equiv\lambda
\end{equation}
and we are looking for the states with $ \lz=2\lambda$. The {\it
W} state is a such state; this is easy to see by setting $a=b=c$
in \eqref {3.cf4}. These substitutions yield
\begin{equation}\label{5.inw}
   \lz(W)=2\lambda(W)=\sqrt{2}|\lh(W)|,
\end{equation}
which shows that \eqref{5.ineq} is indeed a non-strict inequality and gives the least upper bound for the nondiagonal coefficients.

\section{Missed inequality}

In this section we show that another inequality is needed to
specify uniquely the Schmidt coefficients of three-qubit states.
To prove this statement let us assume the converse. Then
\eqref{5.ineq} is a necessary and sufficient condition and GSD
coefficients should satisfy only \eqref{5.ineq} and $\lz\ge|\lh|$.
Consider symmetric states  and put $\lf=\ls=\lt=\lambda$, which
yields $\lz\ge 2\lambda$. Then there exists a state such that
$\lz=|\lh|=2\lambda$, and its GSD is given by
\begin{equation}\label{6.wrongpsi}
   \ket{\psi_{contr}} =\frac{1}{\sqrt{11}} \left(2\ket{000} + \ket{011} + \ket{101} + \ket{110} + 2\ket{111}\right).
\end{equation}
This is a wrong GSD. Indeed,
$$\lz^2(wrong)=\frac{4}{11},$$
but it is shown in Ref. \cite{maxim} that the absolute minimum of
$\lz^2$ over three-qubit pure states is 4/9, and this minimum is
reached at the {\it W} state. Hence no three qubit state exists
for which $\lz^2<4/9$. For the sake of clarity we present the
maximal product overlap and nearest product state for the state
\eqref{6.wrongpsi},
\begin{equation}\label{6.right}
   \lz^2(right)=\frac{14+3\sqrt{2}}{22},\quad \ket{u_1u_2u_3} = (\cos\theta\ket{0}+\sin\theta\ket{1})^{\otimes3},\quad \tan\theta=1+\sqrt{2},
\end{equation}
which can be derived by usual maximization tools.

 This example shows that conditions $\lz\ge 2\lambda$ and $\lz\ge |\lh|$ are insufficient and another relation should exist, and this new relation should give bounds for the last Schmidt coefficient. We know that when all nondiagonal coefficients vanish the upper bound is $|\lh(\max)|=\lz$(known as the GHZ state), and when all nondiagonal elements are maximal given by \eqref{5.inw} the upper bound is $|\lh(\max)|=\lz/\sqrt{2}$ (at the {\it
W} state). Hence for $|\lh|$ there exists an upper bound depending
on the remaining coefficients, and this upper bound gives those
particular bounds at GHZ and {\it W} states, respectively.

 We can derive this upper bound in some simple cases, for instance, when $\ls=\lt=0$ and the state is
 \begin{equation}\label{6.simple}
   \ket{\psi_{simple}}=\lz\ket{000}+\lf\ket{011}+\lh\ket{111},
 \end{equation}
 where $\lh$ is positive as its phase is meaningless in this case.

 The stationarity equations \eqref{2.stf} for the state \eqref{6.simple} have a relevant solution given by
 \begin{equation}\label{6.simsol}
    \ket{u_1}=\frac{\lf\ket{0}+\lh\ket{1}}{\sqrt{\lf^2+\lh^2}},\quad \ket{u_2}=\ket{1}, \quad \ket{u_3}=\ket{1}, \quad \lz^\p=\sqrt{\lf^2+\lh^2}.
 \end{equation}
 From this solution it follows that \eqref{6.simple} is a right decomposition if and only if $\lz\ge\lz^\p$, that is
 \begin{equation}\label{6.simboun}
   \lz^2\,\ge\lf^2+\lh^2, \quad \ls=\lt=0.
 \end{equation}
 This inequality gives the least upper bound for the last Schmidt coefficient when two nondiagonal coefficients vanish. Unfortunately the tools used in this work were unable to find the least upper bound of $|\lh|$ for generic states.

\section{Summary}

The main result of this work is the inequality \eqref{5.ineq}. Its
role is to separate out three-qubit Schmidt coefficients from the
set of four positive and one complex numbers. As explained in the
above section, it is a necessary but not a sufficient condition,
and another inequality should exist to complete the task.

It is likely that the three nondiagonal elements together define
bounds for the last Schmidt coefficients in the missed inequality.
Then the nondiagonal coefficients are not just extra terms in the
GSD, but the ones which can show some important features of
tripartite entanglement unknown so far.

Another application of the derived nonstrict inequality is that it
can give us a hint how we extend Nielsen's protocol or operational
entanglement measures to three-qubit states. For instance, in the
bipartite case the protocol relies on inequalities quadratic on
Schmidt coefficients. In the three-qubit case such a theorem
should include cubic relations as is evident from \eqref{5.ineq}.

\begin{acknowledgments}
This work was supported by the grant ``Innovation technologies
with young scientists-2012.''
\end{acknowledgments}

\end{document}